\documentclass{eas}
\usepackage{graphicx}
\usepackage{natbib}
\def\hho{H$_2$O}

\def\hhhop{H$_3$O$^+$}

\def\pow#1#2{#1$\times$10$^{#2}$}

\def\ltsim{{_<\atop{^\sim}}}

\def\vhel{$V_{\rm helio}$}

\def\kms{km~s$^{-1}$}

\TitreGlobal{Far-infrared observations of the interstellar medium}
\Editors{Editors: C. Kramer, R. Simon et al.}
\begin{document}

\title{Extragalactic \hhhop: Some Consequences} 
\author{Floris van der Tak}\address{SRON, Landleven 12, 9747 AD Groningen, The
  Netherlands; vdtak@sron.nl}
\author{Susanne Aalto}\address{Onsala Space Observatory, Sweden}
\author{Rowin Meijerink}\address{University of California at Berkeley, USA}
\begin{abstract}
  We discuss some implications of our recent detection of extragalactic \hhhop:
  the location of the gas in M82, the origin of energetic radiation in M82, and
  the possible feedback effects of star formation on the cosmic ray flux in galaxies.
\end{abstract}
\maketitle
\section{Introduction}

Last year saw the first detection of the \hhhop\ molecule outside the Galaxy
\citep{vdtak:xgal}. Using the new 16-pixel HARP imaging spectrometer on the
James Clerk Maxwell Telescope, line emission at a rest frequency of 364\,GHz was
detected towards the prototypical starburst galaxy M82 and the prototypical
ultraluminous merger Arp~220. The derived \hhhop\ abundances and \hhhop/\hho\
ratios imply very high ionization rates for the dense molecular gas in the
nuclei of these two galaxies. Chemical models \citep{meijerink:grid} indicate
that the origin of this high ionization is irradiation by X-rays in the case of
Arp~220, whereas for M82, a combination of ultraviolet light and cosmic rays is
needed. The high ionization rates of these galactic nuclei make magnetic
fields more effective in retarding their gravitational collapse and the
formation of stars.  
%
%
The origin of the irradiation in M82 is the evolved starburst
\citep{schreiber:m82}, whereas for Arp~220, an AGN may be needed, as claimed
before by \citet{downes:arp220}.
Naturally, some questions remain, of which this paper discusses a few.

\section{Location of the \hhhop\ in the nucleus of M82}

The velocities and widths of the two components of the \hhhop\ line profile
observed toward M82 may be used to constrain the location of the gas.
Observations of CO emission lines toward the M82 nucleus show a flattened
structure of size $\approx$50$\times$20$''$ (1000$\times$400~pc) with emission
peaks on its north-eastern and south-western ends. The velocity field is well
described by a monotonic gradient along the major axis, so that the structure is
probably a rotating ring or torus. Observations of CO 6--5 with the JCMT
\citep{seaquist:m82} show a third peak in the middle, which is even better
visible in interferometric images of CO 1--0 \citep{walter:m82} and known as the
`central hotspot'.

The simple velocity structure of the rotating ring implies that the velocities
of the \hhhop\ components correspond to unique positions.  In particular, the
velocity of the narrow component of \vhel$\approx$270\,\kms\ corresponds to the
inner edge of the north-eastern peak, which is just covered by the beam of the
\hhhop\ observations. The width of the narrow component agrees with that of the
CO emission from the north-eastern peak.

The velocity of the broad \hhhop\ component of \vhel$\approx$220\,\kms\
corresponds to a position right between the two main CO peaks. Since the width
of this component is much larger than the systematic velocity gradient within
one JCMT beam, the most plausible origin of the broad component is the `central
hotspot' of molecular emission. This hotspot is located close to the dynamical
center of M82 and shows large line widths in several other molecular lines.  The
distribution of \hhhop\ in M82 thus appears to mimic that of other dense gas
tracers such as HCN \citep{mauersberger:n2h+}.

\section{Origins of energetic radiation in M82}

The X-ray luminosity ($L_X$) of M82 on arcminute scales as measured by ROSAT and ASCA is
$\sim$\pow{2}{41}\,erg/s after correction for internal absorption \citep{moran:xrays}.
In the central $<sim$100<,pc of M82, where \hhhop\ has been detected, both the
nuclear non-thermal component and the central thermal component may contribute
to the ionization of the molecular gas; the superwind is too far away and its
luminosity too low for it to play a role.
Observations with \textit{Chandra} and XMM-\textit{Newton} \citep{zezas:xrays}
seem to indicate that the emission in the extended nuclear region is dominated
by high-mass X-ray binaries (HMXBs) and so-called ultraluminous X-ray sources
(ULXs). The X-ray emission from a starburst is dominated by HMXBs, such that
$L_X$ may be used to estimate its star formation rate \citep{grimm:xrays}.
Future spatially and spectrally resolved observations of \hho\ line emission in
M82 may be used to determine the \hhhop/\hho\ ratio for both components of \hhhop\
emission and to characterize the ionization state of the molecular gas in M82 as a
function of position in the nuclear disk.

\section{Effects of star formation on the cosmic-ray flux}

The main astrophysical interest in \hhhop\ is the use of the \hhhop/\hho\
abundance ratio as a probe of the ionization rate of dense molecular gas. 
This rate is an important parameter for the ability to form stars, because it
determines whether magnetic fields may be effective in supporting the cloud
against gravitational collapse.
The ionization rate of dense molecular clouds is dominated by ultraviolet light
and X-rays near their surfaces, and low-energy cosmic rays ($E \ltsim 1$\,GeV) in their interiors.
Foreground absorption often hides these types of radiation from direct
observation, which is why \hhhop\ observations are so useful.

Recently, \cite{pellegrini:m17} proposed a model for the photodissociation
region M17 where the thickness of the atomic gas layer between the ionized and
molecular components is determined by magnetic pressure rather than gas
pressure. The motivation for this model is the unusually strong magnetic field
measured in M17 through the Zeeman effect on the HI 21\,cm line \citep{brogan:m17}.
Similar results were found for the neutral gas in front of the Orion nebula (the
so-called veil) by \cite{abel:orion}. The idea is that soon after the formation
of a star cluster, starlight momentum and gas pressure from the ionized region
compress the surrounding neutral gas, where the magnetic pressure builds up
until it can withstand the external pressure. See also Ferland (this volume).

A consequence of Pellegrini's model would be that the cosmic-ray flux in the gas
surrounding the star cluster is also enhanced, if the cosmic rays are trapped in
the magnetic field. The efficiency of this trapping depends on the geometry of
the magnetic field, in particular whether the field lines are open or
closed. Direct tests of this model are not easy because the Zeeman effect is
difficult to measure. However, an indirect test would be to look for
correlations between the cosmic-ray ionization rates of star-forming regions
with their stellar luminosity.

\begin{figure}
  \centering
  \includegraphics[width=8cm,angle=-90]{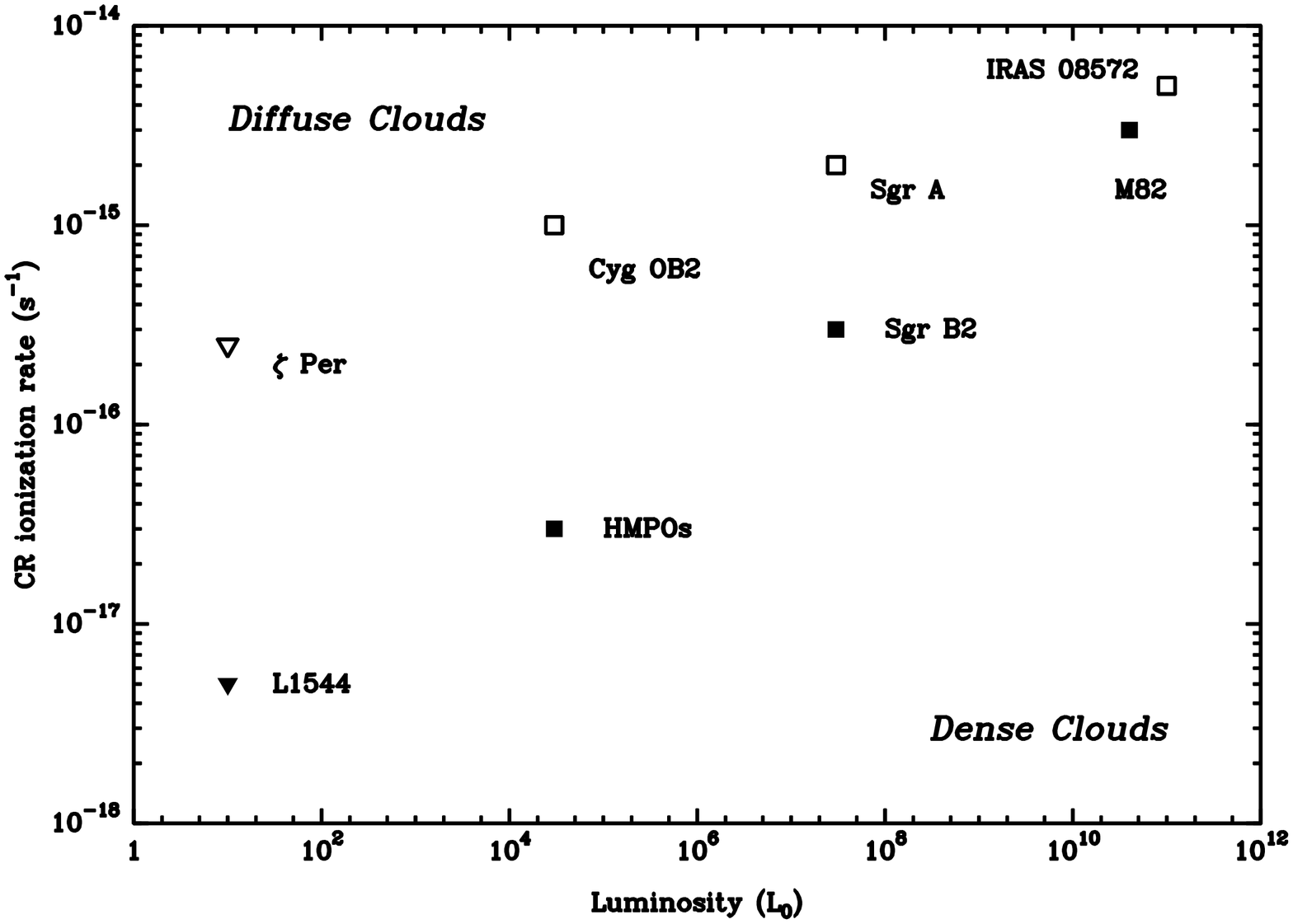} 
  \caption{Relation between cosmic-ray ionization rate and bolometric
    luminosity. Ionization rates are from \citet{caselli:zeta},
    \citet{vdtak:zeta}, \citet{vdtak:sgrb2} and \citet{vdtak:xgal}; luminosities
    are from \cite{vdtak:massive}, \citet{lis:sgrb2} and \citet{spaans:M82}.
    The luminosity of the pre-stellar core L1544 is an upper limit.}
  \label{fig:ion}
\end{figure}

Figure~\ref{fig:ion} shows recent observational estimates of the cosmic-ray
ionization rate in dense molecular clouds plotted against their luminosity. 
Although the estimates are probably only accurate to order of magnitude, a
trend does appear to be visible.
The possible correlation cannot just be a distance effect because the ionization
rates are local quantities derived from the ratio of two molecular abundances.
Thus, star formation may indeed influence the local cosmic-ray flux.

Besides variations with luminosity, the cosmic-ray ionization rate is also known
to differ between diffuse and dense clouds. Ionization rates in diffuse gas are
factors of 3--10 higher than those in dense clouds in the same region which are
exposed to the same incident cosmic-ray flux. 
%
%
This effect is shown by the open symbols in Figure~\ref{fig:ion}, with data from
\citet{lepetit:zPer} for $\zeta$~Per, \citet{mcCall:diffH3+} for Cyg OB2,
\citet{oka:sgra} for Sgr~A and \citet{geballe:xH3+} for IRAS 08572. 
%
%
The offset between the diffuse and the dense clouds must be due to
propagation effects, and scattering of cosmic rays off plasma waves may play a
role \citep{padoan:cr_mhd}, but only at low column densities ($A_V \ltsim 10$).
In the bulk of the clouds, the lifetime $\tau$ of the cosmic rays is probably
limited by energy losses, which scale as $\tau \sim 2\times 10^5 (n_H / 300
\mathrm{cm}^{-3} )^{-1}$\,yr \citep{gabici:gamma}.


\acknowledgements 
The authors thank Gary Ferland, Frank Israel and Sera Markoff for useful discussions.

\setlength{\parsep}{-12pt}
\bibliographystyle{aa}
\bibliography{vdtak}

\end{document}